\input epsf
\documentstyle[prl,aps]{revtex}
\draft
\begin{document}

\twocolumn[
\hsize\textwidth\columnwidth\hsize\csname @twocolumnfalse\endcsname

\date{\today}
\title{Stress-free Spatial Anisotropy in Phase-Ordering}
\author{A. D. Rutenberg \cite{email}}
\address{Theoretical Physics, University of Oxford, Oxford OX1 3NP, United
Kingdom}
\maketitle

\widetext

\begin{abstract}
We find spatial anisotropy in the asymptotic correlations of
two-dimensional Ising models under non-equilibrium
phase-ordering. Anisotropy is
seen for critical and off-critical quenches and both conserved and
non-conserved dynamics. We argue that spatial anisotropy is generic
for scalar systems (including Potts models)
 with an anisotropic surface tension. Correlation functions will not
be universal in these systems
since anisotropy will depend on, e.g., temperature,
microscopic interactions and dynamics, disorder, and frustration.
\end{abstract}
\pacs{05.70.Ln, 64.60.Cn}
\narrowtext
]

Most theoretical, numerical, and experimental
treatments of non-equilibrium phase-ordering
kinetics assume that asymptotic correlations are spatially isotropic  in
stress-free systems quenched from disordered initial conditions into the
ordered phase \cite{Bray94,Maheshwari93}.  Numerical
measurements in simple Ising models have supported this assumption
\cite{Humayun91,Shore92},  despite several qualitative reports to
the contrary \cite{Lacasse93,Marko95}.
It has not been clear whether Potts models \cite{Holm91}
and frustrated 3d Ising models \cite{Shore92}, where anisotropy effects
have been measured, are typical or are special cases.  Indeed, without
a demonstration that anisotropy is {\em expected} asymptotically it has
been possible to discount measured anisotropy as being a transient effect
(see, e.g., \cite{Rao95}).

Naively, it is reasonable to expect anisotropic correlations in phase-ordering.
The 2d Ising model below $T_c$, for example,
is spatially anisotropic at arbitrarily large scales in equilibrium
correlations \cite{McCoy73}, in the resulting surface tension \cite{Avron82},
and hence in the interface dynamics \cite{Siegert90}.  In fact,
we demonstrate below that spatial anisotropy is {\em generic} in scalar
systems for quenches into an ordered phase with an anisotropic surface tension.
Isotropic theories are inadequate for these systems.

For a coarse-grained scalar order parameter, $\phi$, in the continuum limit,
we define an effective free energy in momentum space
$F[\{ \phi \}] = \int d^d k [ (k^2 +D_{\bf k}) \phi_{\bf k} \phi_{-{\bf k}}
+ V_k]$, where
$V_k[\{ \phi \}]$ is the fourier transform of local potential
terms and the differential operator $D_{\bf k}$ includes anisotropic
higher-order gradients. We define an angle
dependent free-energy density $\epsilon({\bf n})$ by restricting the integral
in $F$ to momenta in a given direction ${\bf n}$ and averaging
 (denoted by angle brackets) over the random initial conditions.
To do the restricted momentum integral in $\langle F \rangle$,
 at late enough times for domain walls to be well defined,
we use the anisotropic Porod law for the structure factor
$S({\bf k}) = \langle \phi_{\bf k} \phi_{- {\bf k}} \rangle$
 in general dimension $d$ (generalizing \cite{Onuki92}):
\begin{equation}
\label{EQN:POROD}
	S({\bf k}) \simeq 2^{d+2} \pi^{d-1} A k^{-(d+1)} P({\bf k}/k),
\end{equation}
which holds for $L^{-1} \ll |{\bf k}| \ll \xi({\bf n})^{-1}$,
where $L(t)$ is the characteristic growing
length-scale of the system and $\xi({\bf n})$ is the domain wall width
for domain walls with normal orientation ${\bf n} = {\bf k}/k$.
$A(t) \sim L^{-1}$ is the average area density of domain wall
and $P({\bf n})$ is the angle distribution function of domain
wall orientation [with $P({\bf n})=P(-{\bf n})$ and $\int d{\bf n}
P({\bf n}) =1$].   We find
\begin{equation}
\label{EQN:EANGLE}
	\epsilon({\bf n}) = A \sigma({\bf n}) P({\bf n}),
\end{equation}
where the leading $k^2$ and potential terms make isotropic
contributions to $\sigma({\bf n})$,
while $D_{\bf k}$ makes anisotropic contributions both directly and
through $\xi({\bf n})^{-1}$.
{}From Eqn. (\ref{EQN:EANGLE}), we identify $\sigma({\bf n})$ as the
effective angle dependent surface-tension for domain walls with normal
direction ${\bf n}$.

Now consider a dissipative quench to $T=0$, with no thermal noise.
The dynamics will be given by
$\dot{\phi}_{\bf k} = - \Gamma({\bf k}) \delta F/ \delta \phi_{-\bf k}$, where
the dot indicates a time-derivative.
$\Gamma = {\rm const.}$ corresponds to
non-conserved dynamics and $\Gamma= \Gamma_2 k^2$ to conserved dynamics
\cite{leading}.
The time rate of change of the energy-density is then simply \cite{Bray94b}
\begin{eqnarray}
\label{EQN:ECHANGE}
	\dot{\epsilon}({\bf n}) &=& \int dk k^{d-1} \langle \delta
F/\delta \phi_{\bf k} \dot{\phi}_{\bf k} \rangle \nonumber \\
	&=& - \int dk k^{d-1} \Gamma^{-1}
\langle \dot{\phi}_{\bf k} \dot{\phi}_{-{\bf k}} \rangle,
\end{eqnarray}
where ${\bf k} \equiv k {\bf n}$. We have
used the dynamics of $\phi$ to replace the functional derivative
of the free-energy so that the resulting expression for
$\dot{\epsilon}({\bf n})$ has no explicit dependence on $F[\{ \phi \}]$.

Under the assumption of isotropic correlations, we generically obtain
different anisotropies in Eqns. (\ref{EQN:EANGLE}) and
(\ref{EQN:ECHANGE}) at arbitrarily late times --- a
contradiction since $\dot{\epsilon} \equiv
\partial_t \epsilon$.  Apart from possible anisotropies
in the correlations, the
anisotropy of $\epsilon({\bf n})$ in Eqn. (\ref{EQN:EANGLE}) is
determined by the statics, through $\sigma({\bf n})$. On the other
hand the anisotropy of $\dot{\epsilon}({\bf n})$ in Eqn. (\ref{EQN:ECHANGE}) is
determined by the dynamics, through $\Gamma({\bf k})$, in addition to
possible contributions by the effective UV cutoff $\xi({\bf n})^{-1}$.
Since $\sigma$ and $\Gamma$ are independent,
the anisotropies will not generally be
equal unless anisotropic correlations make up the difference.
For the general case, anisotropy in $\sigma({\bf n})$ implies anisotropy
in $S({\bf k})$ {\em even in the scaling limit} \cite{scalinglimit}.

The renormalization-group (RG) approach to phase-ordering  \cite{Bray89} is
easily generalized to include anisotropy. The only change is to note
that any anisotropy of either $F[\{ \phi \}]$ or
$\Gamma({\bf k})$ will be renormalized by microscopic details.
(An illustration of this renormalization
is the temperature dependence of the effective surface tension
\cite{Avron82}.) The demonstration that thermal
noise will be asymptotically irrelevant for quenches to below $T_c$
will still apply, with the caveat that the effective $T=0$ dynamics will
include the surface-tension at the quench temperature.  We
then apply our above argument that predicts anisotropy with noise free
dynamics. As a result, we expect anisotropy for all scalar quenches
below $T_c$ \cite{Tc}. Anisotropy may
be renormalized by temperature, disorder, the
details of the local interactions in the system (such as frustration),
and even by global conservation
laws that are ``irrelevant'' \cite{Bray89} in terms of growth laws.
Anisotropy may also depend on the details of the microscopic dynamics.

In principle, we could try to choose the anisotropy of
$\Gamma({\bf k})$ to allow isotropic correlations despite an
anisotropic $\sigma({\bf n})$. This
fine-tuning of the dynamics ($\Gamma$) with respect to the statics
($\sigma$) will not generically occur. The RG approach \cite{Bray89} shows that
$\Gamma({\bf k})$ will only be renormalized analytically, i.e. anisotropy
will only enter at $O(k^4)$
and higher.  For conserved dynamics, these contributions
are subdominant in Eqn. (\ref{EQN:ECHANGE}) since
 the integral converges in the UV \cite{Bray94b}.  Thus neither
the anisotropy of $\Gamma({\bf k})$ nor the
anisotropy of the core scale $\xi({\bf n})$ will
affect $\dot{\epsilon}({\bf n})$ through Eqn. (\ref{EQN:ECHANGE}).
For conserved dynamics, even fine tuning of $\Gamma$ cannot eliminate
anisotropic correlations.
With non-conserved dynamics, the  UV regime
dominates the energy-dissipation integral (\ref{EQN:ECHANGE})
\cite{Bray94b} and both  $\Gamma({\bf k})$ and $\xi({\bf n})$ make
anisotropic  contributions to $\dot{\epsilon}({\bf n})$.
In principle the anisotropy of $\Gamma({\bf k})$ could be renormalized to
compensate $\sigma({\bf n})$ --- allowing isotropic correlations within
our argument. However, we find anisotropy numerically for $T>0$ even for
non-conserved dynamics. We deduce that
fine tuning of the kinetic prefactor does not occur.

For a system defined on a lattice, anisotropies will be present
in the surface tension at $T=0$, because
lattice interactions are not rotationally invariant.
Our argument then implies
anisotropic correlations in the scaling limit. Anisotropy is only
implied in some of the
correlation functions and not necessarily in two-point correlations.
In practice we find, in all of our numerical studies,
significant anisotropy effects in the two-point correlations.

For the remainder of this letter, we
explore 2d Ising models with nearest-neighbor interactions on a square
lattice under quenches from random
initial conditions. We find anisotropic correlations quite generally ---
for a variety of temperatures below $T_C$,
of initial magnetizations, and
for all of globally conserved, non-conserved, and locally conserved dynamics.
These anisotropies do not decrease at late times, as would be expected
for transient effects introduced by the dynamics at earlier times.

\begin{figure}[t]
\begin{center}
\mbox{
\epsfxsize=3.0in
\epsfbox{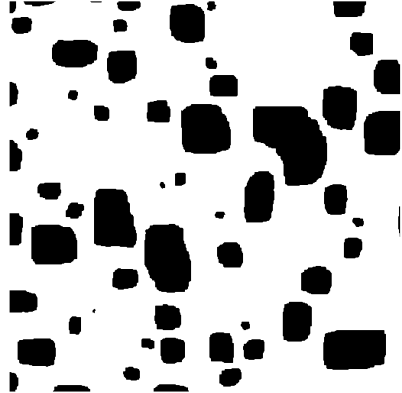}}
\end{center}
\caption{A $512^2$ region of a globally conserved
$T/T_c=0.2$ quench with $\langle \phi \rangle =0.4$ ($t=513$ MCS).
Lattice directions, here and in the next figure, are vertical and horizontal.
\label{FIG:configuration}}
\end{figure}

We measure the normalized correlations
$C({\bf r},t) = (\langle  \phi({\bf r}) \phi(0) \rangle - \langle
\phi \rangle^2)/(\langle \phi(0^+) \phi(0) \rangle - \langle \phi \rangle^2)$,
which ranges from $1$ at short distances to $0$ at infinity
\cite{scalinglimit}.  The anisotropic length-scale $L({\bf n},t)$
of a system is defined by the scale in direction $\bf n$ at
which $C=0.5$  for non-conserved dynamics, and by the first zero of $C$
for conserved
dynamics.  We scale correlations in all directions by the length-scale
in the diagonal direction. A natural measure of anisotropy is
$\chi = (L_{max}/L_{min}-1)/(\sqrt{2}-1)$, where $L_{max}$ is the
maximum length-scale at a given time, $L_{min}$ is the minimum, and
the normalization makes $\chi$ run from $0$ (circle) to $1$ (square) for
convex contours of $C({\bf r})$.

\begin{figure}[t]
\begin{center}
\mbox{
\epsfxsize=2.8in
\epsfbox{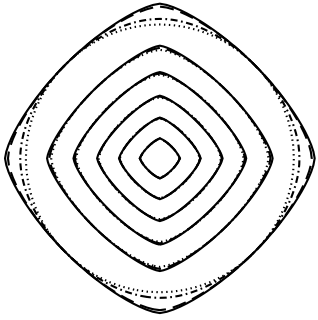}}
\end{center}
\caption{Anisotropic contours of scaled correlations $C({\bf r}/L) = 0.9$,
$0.8$, $0.7$, $0.6$, and $0.5$ (from the center) for
an off-critical quench to $T=0$ with $\langle \phi \rangle = 0.4$ and
globally conserved Creutz dynamics. The
times are $t=513$ MCS (dotted), $1025$ (dashed), and $2049$ (solid). Also
shown, scaled by $1.4$ for clarity, are the $C=0.5$ contours of quenches to
$T/T_c = 0$, $0.2$, $0.4$, and $0.9$ (solid, dashed, dot-dashed, and
dotted lines, respectively) at $t=1025$ MCS. The length-scale $L$ is such
that, along the diagonal direction, $C(L)=0.5$.
\label{FIG:contour}}
\end{figure}

\begin{figure}[t]
\begin{center}
\mbox{
\epsfxsize=2.8in
\epsfbox{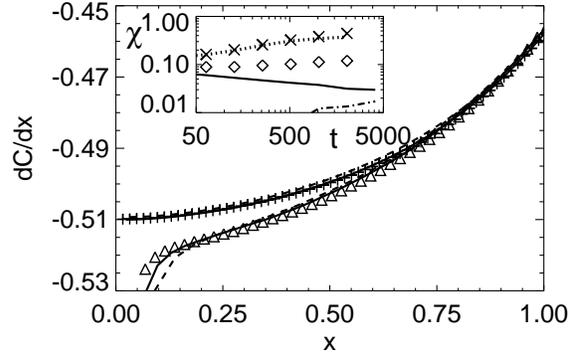}}
\end{center}
\caption{$dC/dx$ vs
scaled distances, $x$, for a non-conserved critical quench to
$T=0$.  The plusses indicate correlations in the lattice
axis direction, while the triangles indicate correlations along lattice
diagonals (at $t=4097$ MCS).
Solid and dashed lines indicate corresponding correlations
at $t=2049$ and $1025$ MCS respectively. In the inset is the anisotropy
measure $\chi$ vs. $t$ for $T/T_c=0$, $0.2$, and $0.4$ globally-conserved
quenches from Fig. 2 (crosses, dotted lines, and diamonds,
respectively). At the bottom of the inset are the bare and corrected $\chi$
(solid and dot-dashed lines, respectively) for the non-conserved quench
of the main figure.
\label{FIG:smallx}}
\end{figure}

We first consider off-critical quenches
with a global conservation law to prevent the magnetization from
saturating. We couple the system to a Creutz spin reservoir of size $2$
\cite{Creutz83}: each randomly chosen spin
is updated by a Metropolis algorithm, subject to an additional
microcanonical constraint that any spin change ($\pm 2$) fits in the spin
reservoir. We study size $1024^2$ systems with
$\langle \phi \rangle = 0.4$.  A snapshot from a quench to $T=0.2 T_c$ in
Fig.  \ref{FIG:configuration} illustrates the strong anisotropy even
above the roughening transition \cite{rough}.
We show some contour plots of the scaled correlations
in Fig. \ref{FIG:contour}.
It is clear that the anisotropies are not limited to the small $r$ regime.
The anisotropy is increasing at late times
(see inset of Fig. \ref{FIG:smallx}).
In the same regime, the spherically averaged correlations scale well.
The latest anisotropies,
at $t=2049$ MCS,  before finite-size effects entered were
$\chi=0.45$ ($T=0$), $0.38$ ($T=0.2 T_c$), and $0.12$ ($T=0.4 T_c$).
[Statistical error bars, with at least $30$ samples in each case, are less than
$\pm 0.001$.]

We also studied non-conserved critical quenches.
With heat-bath dynamics and a sublattice update,
late times in large lattices could be reached. Even so, the asymmetry
remained small. For lattices of size $2048^2$, and a quench to $T=0$, we
show $dC/dx$ vs. $x$ ($x=r/L$), along the diagonal and lattice directions,
in Fig. \ref{FIG:smallx} \cite{Tomita84}.
We find $\chi = 0.03$ at the latest time ($t=4097$ MCS, $74$ samples,
statistical error $\pm 0.0003$). The
constant but orientation dependent domain wall width, $\xi({\bf n})$,
evident in the sharp
downturn of $dC/dx$ near $x=0$ in the  diagonal correlations,
increases $\chi$ by $O(1/L)$.  This
is significant for small anisotropies and early times. Directly
subtracting this $O(1/L)$ contribution leads to $\chi$ slowly increasing
with time (inset of Fig. \ref{FIG:smallx}), with a corrected latest value
$\chi = 0.02$.

\begin{figure}[t]
\begin{center}
\mbox{
\epsfxsize=3.0in
\epsfbox{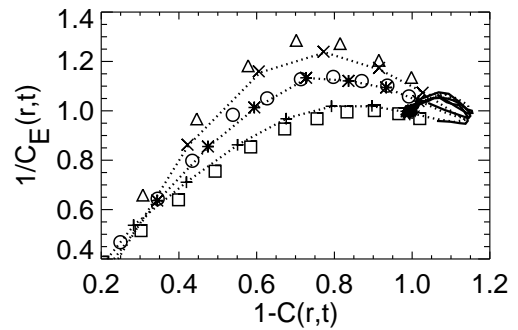}}
\end{center}
\caption{Two-point correlations vs energy-density
energy-density correlations for a conserved quench to $T/T_c = 0.4$ with
$\langle \phi \rangle =0.4$ (size $256^2$).
The correlations are spherically averaged (stars and circles),
along the axis (plusses and squares), and along the
diagonal (crosses and triangles). For clarity, solid lines have been used
after the first zero of $C$.
Times are $2.6 \times 10^5$ and $1.0 \times 10^6$ MCS, respectively.
Dotted lines show data from size $128^2$ systems ($36$ samples) at the
earlier time.
\label{FIG:conserved}}
\end{figure}

We have also simulated conserved 2d Ising systems with nearest-neighbor
Kawasaki exchange dynamics and a Metropolis update. At low enough temperatures
for the anisotropy of $\sigma({\bf n})$ to be visible, the
activated dynamics slows the
simulations considerably. We explored size $256^2$ systems, with
$\langle \phi \rangle = 0.4$ and $T = 0.4 T_c$, up to times
$t=10^6$ MCS ($10$ samples). The length scales achieved ($L \lesssim 12$)
are so small that $\chi \approx 0$ within numerical accuracy,  so in
Fig. \ref{FIG:conserved} we plot $C({\bf r})$ against
the energy-energy correlation function
$C_E({\bf r}) \equiv \langle  E({\bf r}) E(0) \rangle/ \langle E
\rangle^2-1$, where $E({\bf r})$ is the number of broken bonds at
site ${\bf r}$ minus the equilibrium bulk average.
This shows a significant and
increasing difference between correlations in the lattice and diagonal
directions.

In summary of our numerical results, we find anisotropic correlations in
various quenched 2d Ising models.  Anisotropy increases with decreasing
temperature and for increasing net magnetization.  Anisotropy effects
are always slowly {\em increasing} at the latest times of our
simulations.  In all of our simulations, the spherically
averaged correlations scale reasonably well while the anisotropy is still
evolving. To study the non-zero asymptotic anisotropy in these
systems, some sort of acceleration method is needed  (see,
e.g., \cite{Marko95}) --- though in general the anisotropy will depend on the
numerical algorithm used.

In disordered \cite{Bray94,Fisher88} and frustrated \cite{Shore92,Rao95}
models, it has been argued that scaling functions will be ``universal'' ---
identical to those of the Ising systems that
we have studied in this letter. [While logarithmic growth is seen,
it is thought to come from $L$
dependence in the kinetic prefactor $\Gamma$ ---
so that scaled correlations are unaffected.]  However, anisotropy should be
renormalized by disorder and frustration, so we do
{\em not} expect scaling function universality to hold in these systems.
In frustrated 2d and 3d Ising models, the
fairly large anisotropy seen numerically \cite{Shore92,Rao95} should
remain at arbitrarily late-times \cite{rough}.
Hopefully, in experimental random-field systems (e.g. \cite{Feng95}),
anisotropy can be measured directly.

We can generalize our argument around Eqns.
(\ref{EQN:POROD})-(\ref{EQN:ECHANGE}) to systems
with other types of singular defects.
Using vector $O(n)$ order-parameters, and a
generalized Porod's law $S({\bf k}) = D({\bf k}/k) L^{-n} k^{-(d+n)}$
\cite{Bray94}, we find that the anisotropic contribution to the energy
density is asymptotically negligible for systems without domain walls
\cite{XY}.   However,
other systems with dissipative dynamics in which domain walls dominate the
asymptotic energetics will be anisotropic if the surface tension is
anisotropic,
e.g. Potts models (see \cite{Holm91}).

The growth laws of the characteristic length scale $L(t)$ will
remain independent of any anisotropies present, as long as dynamical
scaling is maintained.
This follows from the Energy-Scaling approach \cite{Bray94b} since
anisotropy does not
change the scaling properties of the energy or the rate of energy
dissipation. We would be surprised if the anisotropy affected the
dynamical scaling of the correlations (see however \cite{Marko95}),
though the scaling regime seems to be pushed to much later times as the
anisotropy slowly develops.
It remains an open question whether non-zero anisotropies have
implications beyond the scaled correlations, such as in autocorrelation
exponents.

In practice, isotropic theories have worked
fairly well for spherically averaged correlations.  Certainly, lattices,
interactions, and dynamics can be chosen to minimize anisotropies.
This would be desirable, for instance, in lattice simulations of
isotropic fluid or polymer systems.
However, the language of an {\em isotropic} zero-temperature
phase-ordering fixed point is inappropriate for a
scalar system with an anisotropic surface tension.

In summary, we expect anisotropy for
any scalar lattice system quenched to below $T_c$, including
disordered and/or frustrated systems.
We expect the anisotropy to depend on the details of the system.
Scaled correlation functions of anisotropic
systems will {\em not} be universal.

I thank J. Cardy for discussions. This work was supported by EPSRC
grant GR/J78044.


\begin{references}
\bibitem{leading} We assume isotropy of the leading $k^2$ terms in $F$ and
$\Gamma$.
\bibitem{Tc} At $T_c$, where the surface tension is asymptotically isotropic
\cite{Avron82}, we expect isotropic correlations in the scaling limit.
\bibitem{rough} The presence of an equilibrium roughening transition  does not
affect our argument for the existence of anisotropy. It may, however,
change the nature of the anisotropy.
\bibitem{XY} For XY systems, the anisotropic contribution to the energy
is only down by a factor of $\log{L}$.
\end{references}
\end{document}